\documentclass[floats,floatfix,showpacs,amssymb,prd,twocolumn,superscriptaddress,nofootinbib]{revtex4-1}
\usepackage[dvips, pdftex]{graphicx}
\usepackage{epstopdf}
\usepackage{bm}
\usepackage{longtable}
\usepackage[usenames,dvipsnames]{xcolor}
\usepackage{color}
\usepackage[breaklinks]{hyperref}
\usepackage{amsfonts,amsmath,amssymb,mathrsfs}

\def\be{\begin{equation}}
\def\ee{\end{equation}}
\def\beq{\begin{eqnarray}}
\def\eeq{\end{eqnarray}}
\def\nn{\nonumber}

\begin{document}
\title{NUMERICAL SIMULATION OF COSMIC STRING MOTION IN PERES BACKGROUND}

\author{\firstname{S.~N.}~\surname{Roshchupkin}}
\email{sstringsr@gmail.com}
\affiliation{Taurida National V.~I.~Vernadsky University, Simferopol, Crimea, Ukraine}
\author{\firstname{A.~V.}~\surname{Zhovtan}}
\email{alex\_ph@fastmail.fm}
\affiliation{Crimean Astrophysical Observatory, Nauchnuy, Crimea, Ukraine}

\begin{abstract}
The dynamics of closed cosmic strings in Peres background, using finite-difference method, has been considered. It is shown that motion of the strings is very complicated and its perform complex oscillations and deformed to form loops and cusps.
\end{abstract}

\maketitle

\section{Introduction}

Spontaneous symmetry breaking in gauge theories leads to phase transitions in the early Universe \cite{Peebles}. In this period can appear topological structures
in the Universe like strings, vacuum domain walls and monopoles. One of these topological defects that has drawn more attention is the cosmic string. The cosmic strings are linear topological defects. Only infinitely long and closed loop strings can exist. An infinitely long cosmic string is a static cylindrically symmetric configuration of self-interacting scalar field minimally coupled to a $U(1)$ gauge field.

We have two general type of strings. Those which arise in a phase transition in which a gauge symmetry undergoes spontaneous symmetry breaking is called local or gauge strings, and those which we call the global strings, arise as a result of spontaneous breaking of global symmetry. Strings that have cosmological interest are the local strings.

Local strings have a linear energy density $\mu \sim \eta^2$, where $\eta$ is the vacuum expectation value acquired by the gauge field in the phase transition.

Strings of astrophysical relevance were formed during phase transitions at the grand-unified-theory (GUT) scale ($10^{15}~ GeV$). They have linear energy
density $\mu$ of the order $10^{-6}$. This small value of $\mu$ could justify the weak-field approximation used by Vilenkin. Vilenkin \cite{Vilenkin} using the weak-field approximation obtained that the geometry around a cosmic string is conical, with the deficit angle given by $\delta\phi=8\pi\mu$, where $\mu$ is the linear energy density of the string.

Circular strings in curved backgrounds have been systematically studied \cite{de_Vega1, de_Vega2, Larsen1, Larsen2}, showing interesting deviations from their behavior in flat space-times. Since the equations are nonlinear, it is often quite difficult to obtain exact classical solutions in a variety of curved backgrounds \cite{Anderson}. In papers \cite{Larsen3, Zhovtan} it was shown that the strings equations in the Schwarzschild black hole and Lorentz wormhole space-times are actually non-integrable and exhibit chaotic behavior. It means that it is only possible to find the exact evolution for some special configurations or perform some numerical calculations \cite{Lelyakov}.

The paper is organized as follows. In section 2 we present cosmic string equations of motion. The numerical method and results of numerical simulation for cosmic
strings in Peres background is presented in section 3. In section 4 we discuss our results.

\section{Cosmic string in the curved background}

The Nambu-Goto string action in a curved space-time can be presented in the form \cite{Zheltukhin}
\beq
S=S_0+S_1&=&\int\,d\tau d\sigma \left[\frac{det(\partial_{\mu}x^M G_{MN}(x)\partial_{\nu}x^N)}{E(\tau,\sigma)}\right.\nn\\
&&\left.-\frac{1}{(\alpha')^2} E(\tau,\sigma)\right]\,,
\eeq
where $E(\tau,\sigma)$ is an auxiliary world-sheet density, $M,N,\ldots=0,1,\ldots,D-1$; $\mu,\nu=0,1$ and $\partial_0\equiv\partial/\partial\tau$,
$\partial_1\equiv\partial/\partial\sigma$. The equation of motion for $E(\tau,\sigma)$ produced by Eq.(1) is
\begin{eqnarray}
                 E&=&\alpha'\sqrt{-det~g_{\mu\nu}},\\
g_{\mu\nu}&=&\partial_{\mu}x^M G_{MN}(x)\partial_{\nu}x^N.
\end{eqnarray}
The substitution of $E(\tau,\sigma)$ from Eq.(2) into functional (1) transforms the latter into the Nambu-Goto representation
\begin{equation}
S=-\frac{2}{\alpha'}\int{d\tau d\sigma \sqrt{-det~g_{\mu\nu}}}.
\end{equation}
Thus, the representations (1) and (4) for the string action (1) are classically equivalent. Unlike the representation (4), the representation (1) includes the string
tension parameter $1/\alpha'$ as a constant at an additive world-sheet "cosmological" term playing the role of the potential energy. This term may be considered
as a perturbative addition for the case of a weak tension \cite{Zheltukhin}.

We are to considered a dimensional parameter $\gamma$ or some combination of the parameters defining the metric of the curved space, where the string moves.
Without loss of generality one can put that $\gamma$ has the dimension of $L^2$ ($\hbar=c=1$). Then the value of the dimensionless combination
\begin{equation}
\varepsilon=\frac{\gamma}{\alpha'}
\end{equation}
can be considered as a parameter characterizing the power of string tension. In the gauge
\begin{equation}
E(\tau,\sigma)=-\gamma(x_{,\sigma}^{M}G_{MN}(x)x_{,\sigma}^{N})
\end{equation}
accompanied by the ortho--normality condition
\begin{equation}
x_{,\tau}^{M}G_{MN}(x)x_{,\sigma}^{N}=0,
\end{equation}
where $x_{,\tau}^{M}\equiv\partial x^{M}/\partial\tau$, $x_{,\sigma}^{M}\equiv\partial x^{M}/\partial\sigma$, the variational Euler--Lagrange equations of motion generated by Eq.(1) acquire the form
\begin{equation}
x_{,\tau\tau}^{M}-\varepsilon^2 x_{,\sigma\sigma}^{M}+\Gamma_{PQ}^{M}(G)[x_{,\tau}^{P}x_{,\tau}^{Q}-\varepsilon^2 x_{,\sigma}^{P}x_{,\sigma}^{Q}]=0,
\end{equation}
and they contain the dimensionless parameter $\varepsilon$. This parameter appears in another string constraint
\begin{equation}
x_{,\tau}^{M}G_{MN}(x)x_{,\tau}^{N}+\varepsilon^2 x_{,\sigma}^{M}G_{MN}(x)x_{,\sigma}^{N}=0,
\end{equation}
which is additional to constraint (7). For $\varepsilon=0$ we have the null string. From the above we can see that for the null strings we have the null geodesic equations supplemented by the constraint (7), which ensures that each point of a tensile string (null string) propagates in the direction perpendicular to the string. Thus, knowing the null geodesics in a background space-time would naturally lead to null string configurations provided all the constraints are satisfied.

\section{Numerical simulation of string motion in Peres background}

Let us now discuss the cosmic string propagation in the Peres space-time \cite{Peres}. The metric for such space-time is represented as
\begin{equation}
ds^2=2dtdx-dy^2-dz^2+s(t,y,z)dt^2.
\end{equation}
From Einstein-Hilbert equations we have two different solutions:
\begin{eqnarray}
ds^2=2dtdx-dy^2-dz^2+b(y^2+z^2)dt^2,\\
ds^2=2dtdx-dy^2-dz^2+b(y^2-z^2)dt^2,
\end{eqnarray}
where $b$ is a constant \cite{Ivanov}. Eq.(11) describes radiation field with an isotropic energy-momentum tensor. In the second case (see Eq.(12)) we have space-time which describes strong gravitational waves.

The initial boundary value problem for the relativistic string is as follows. We must find a solution $x^{M}(\tau,\sigma)$ of Eq.(8) that is doubly differentiable in the domain
\begin{equation}
\Omega=\{ (\tau,\sigma): 0\leqslant\tau\leqslant\tau_{max}, 0\leqslant\sigma<2\pi \},
\end{equation}
is continuously differentiable on the boundary $\Omega$, and satisfies constraints (7), (9), the periodicity conditions $x^{M}(\tau,\sigma)=x^{M}(\tau,\sigma+2\pi)$, and two given initial conditions, i.e., the initial string location
\begin{equation}
x^{M}(0,\sigma)=\rho^{M}(\sigma)
\end{equation}
and the initial velocity of string points
\begin{equation}
\dot{x}^{M}(0,\sigma)=v^{M}(\sigma).
\end{equation}
Few results on global solutions of the Cauchy problem for nonlinear hyperbolic systems are known. Equations (7)--(9) describing the strings dynamics pertain to such mathematical models in physics whose theoretical exploration is in the initial stage.

There are two main problems in the theory of numerical calculations. First, one must construct discrete approximations for equations and investigate the a priori
characteristics of the quality of these approximations, which consists primarily in studying approximation inaccuracy, stability properties of approximations, and the
related accuracy of the obtained difference scheme. Second, one must solve difference equations by direct or iteration methods chosen for reasons of economy of
computational time for the calculation algorithm.

We now turn to constructing the difference scheme. For this, we introduce a rectangular lattice in the domain of the world-sheet variables $\tau$ and $\sigma$:
\beq
\tau_{l}=l\Delta\tau,\; l&=&0,\ldots,L,\; \sigma_{m}=m\Delta\sigma,\;\nn\\ 
&&m=0,\ldots,M,
\eeq
where $\Delta\tau=\tau_{max}/L$, $\Delta\sigma=2\pi/(M+1)$, and $L$ and $M$ are the numbers of sites for the respective variables $\tau$ and $\sigma$.
In constructing the difference scheme, we use a method of difference approximation with which we can easily elaborate the scheme of the first- and second-order
approximations on the rectangular lattice for equations with continuous and sufficiently smooth coefficients. We prefer implicit schemes because they converge
better than explicit schemes. We take the template depicted in Fig.~\eqref{fig1} and compose a scheme with the following weights for space derivatives in different layers:
\begin{figure}[h]
\includegraphics[scale=1]{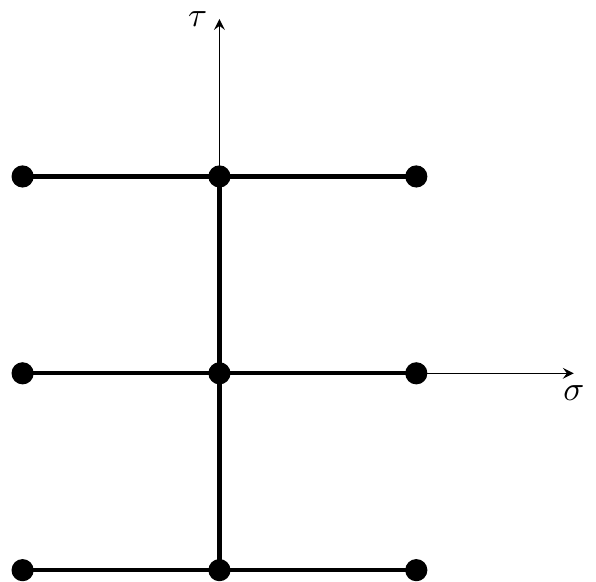}
\caption{Template for the difference scheme.}
\label{fig1}
\end{figure}
\begin{figure*}[htb]
\begin{center}
\begin{tabular}{cc}
\includegraphics[width=7.9cm]{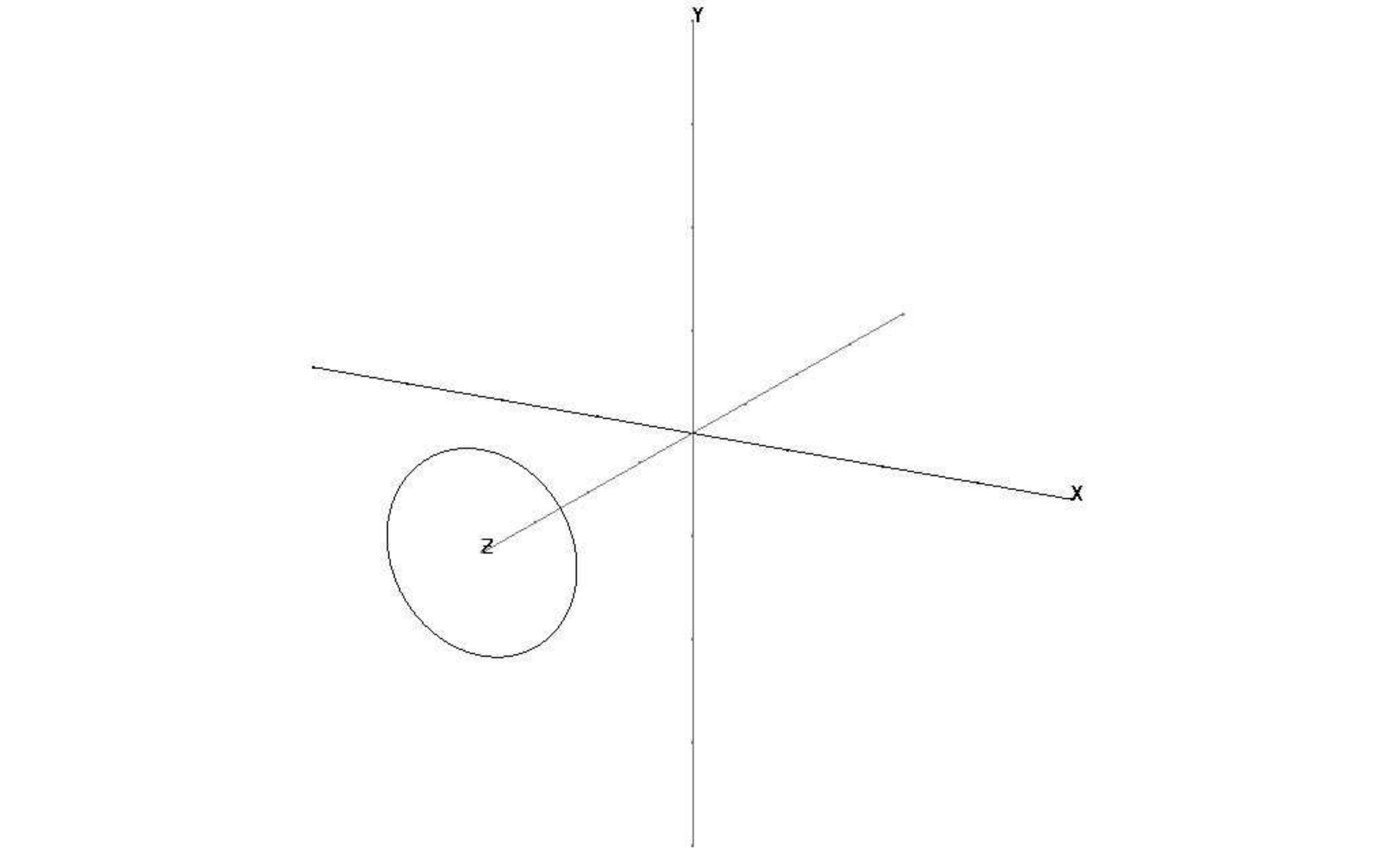}&
\mbox{$\tau=0~~~~~~~~\tau=1$}
\includegraphics[width=7.9cm]{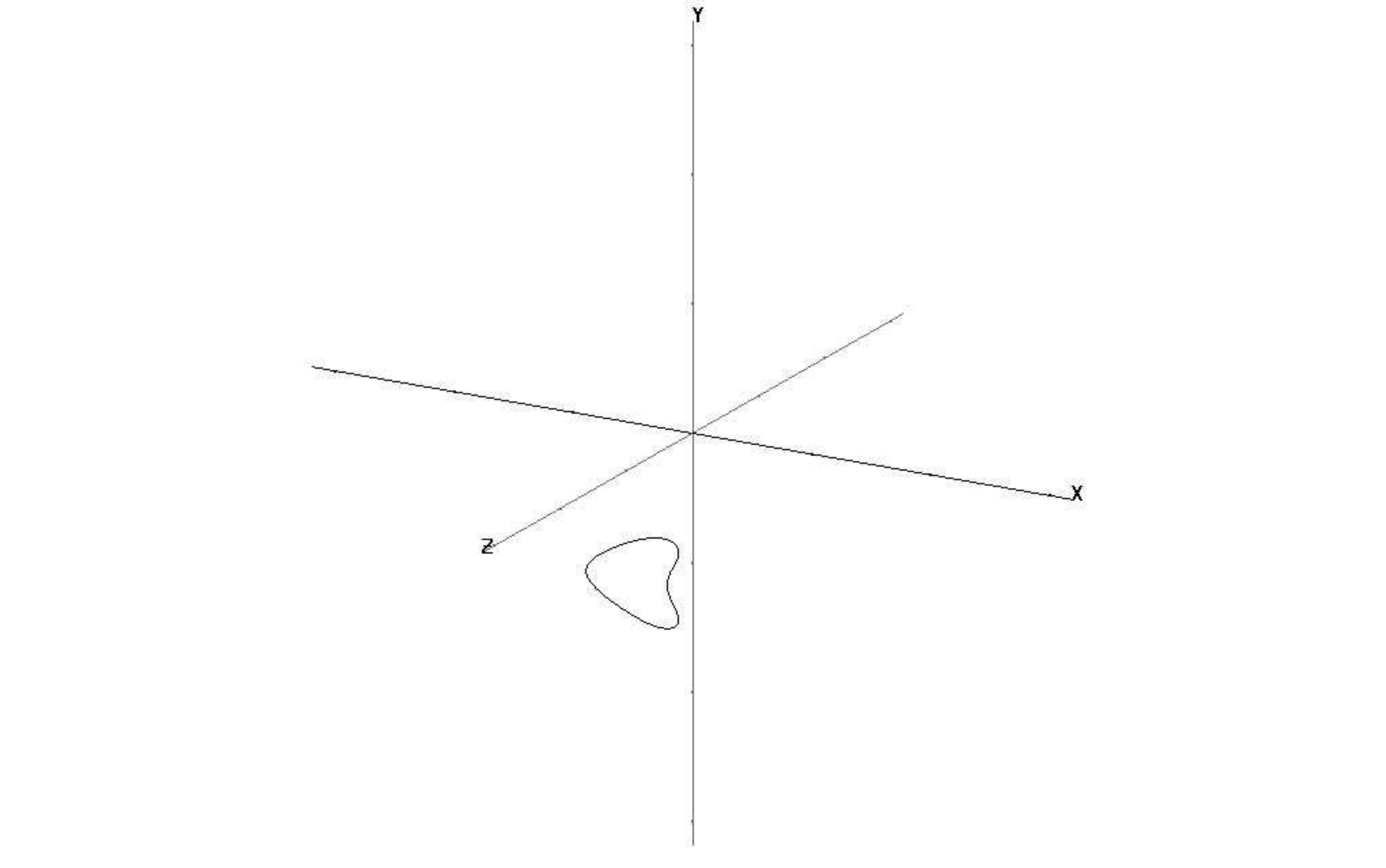}
\end{tabular}
\begin{tabular}{cc}
\includegraphics[width=7.9cm]{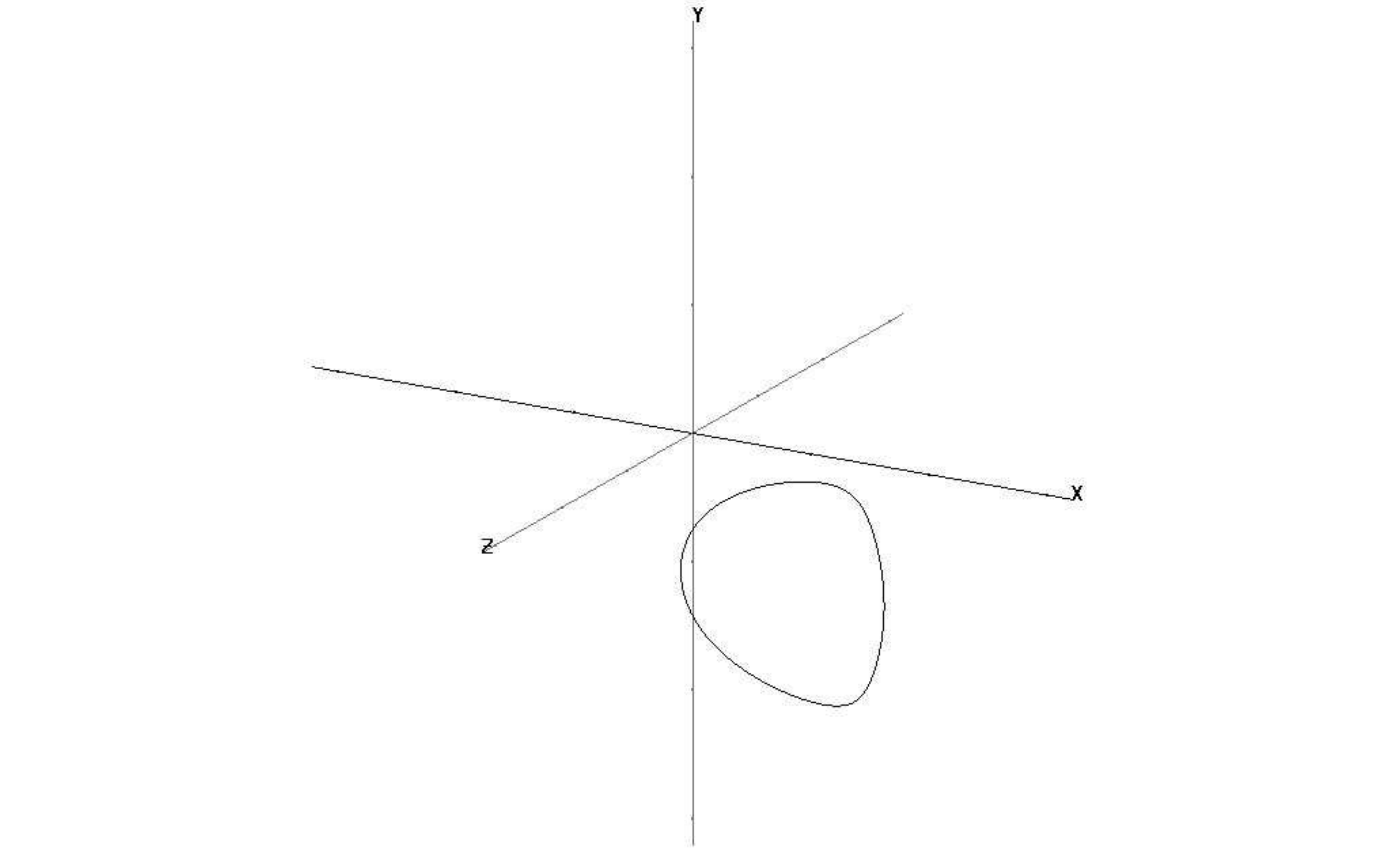}&
\mbox{$\tau=2~~~~~~~~\tau=3$}
\includegraphics[width=7.9cm]{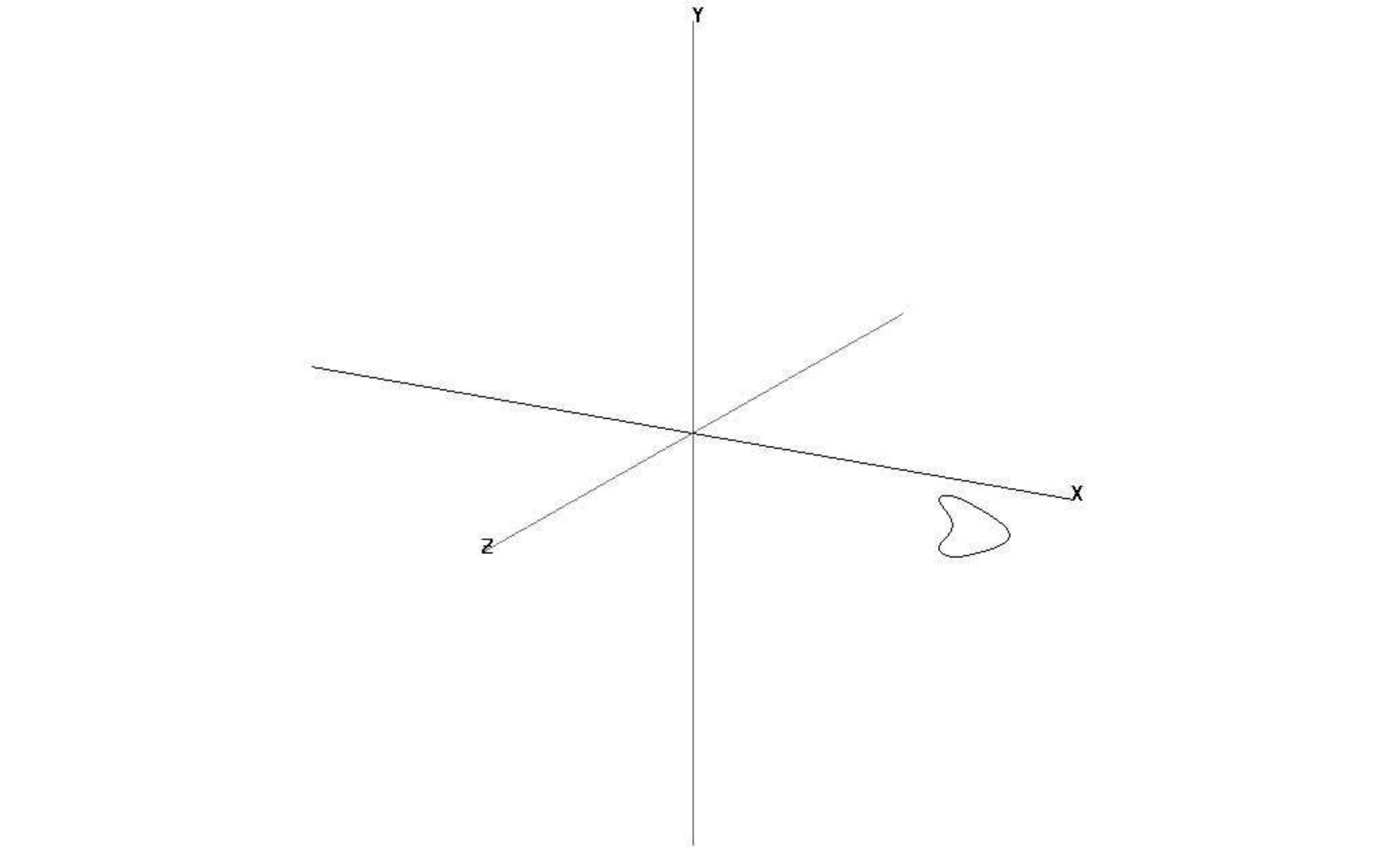}
\end{tabular}
\begin{tabular}{cc}
\includegraphics[width=7.9cm]{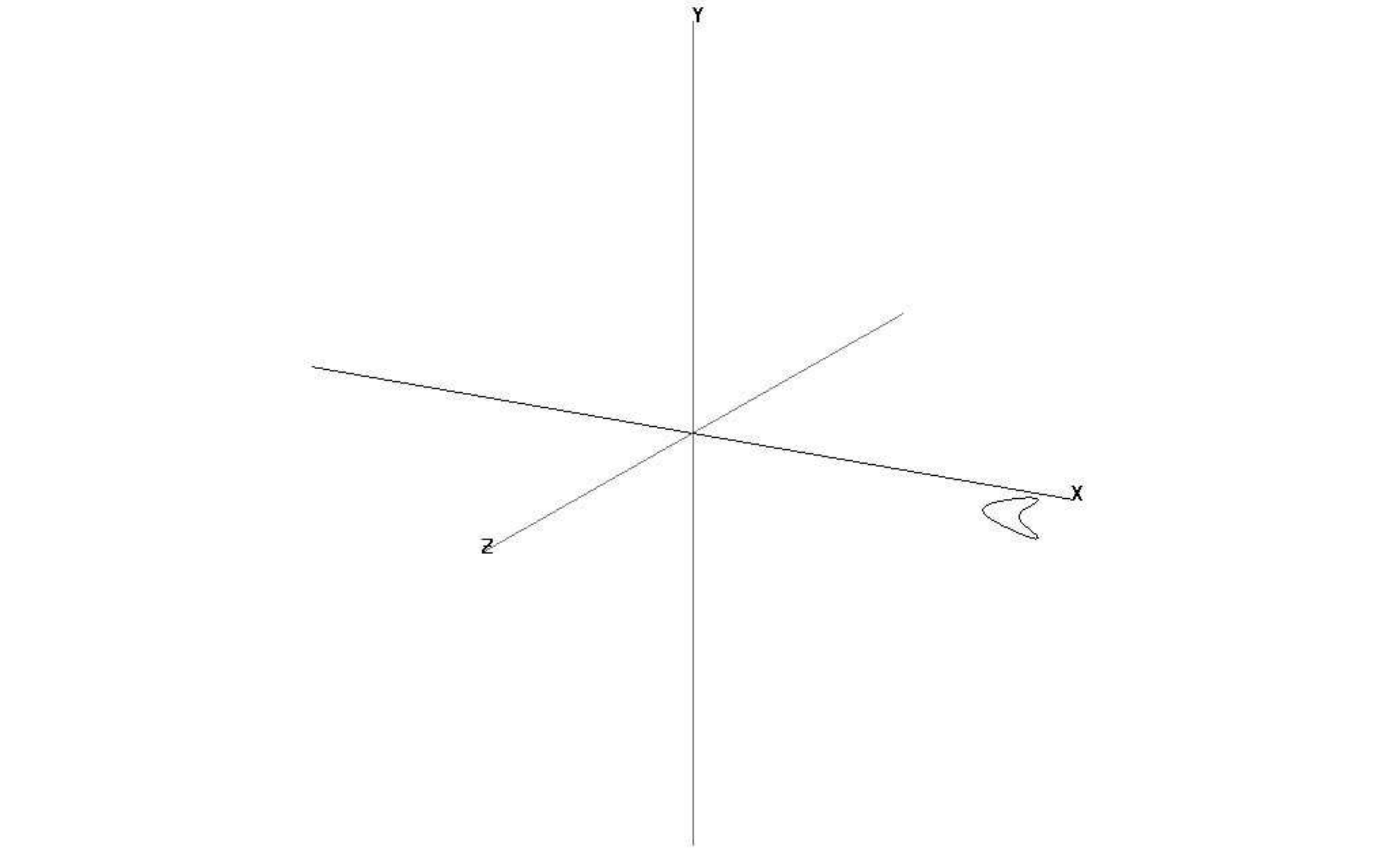}&
\mbox{$\tau=4~~~~~~~~\tau=5$}
\includegraphics[width=7.9cm]{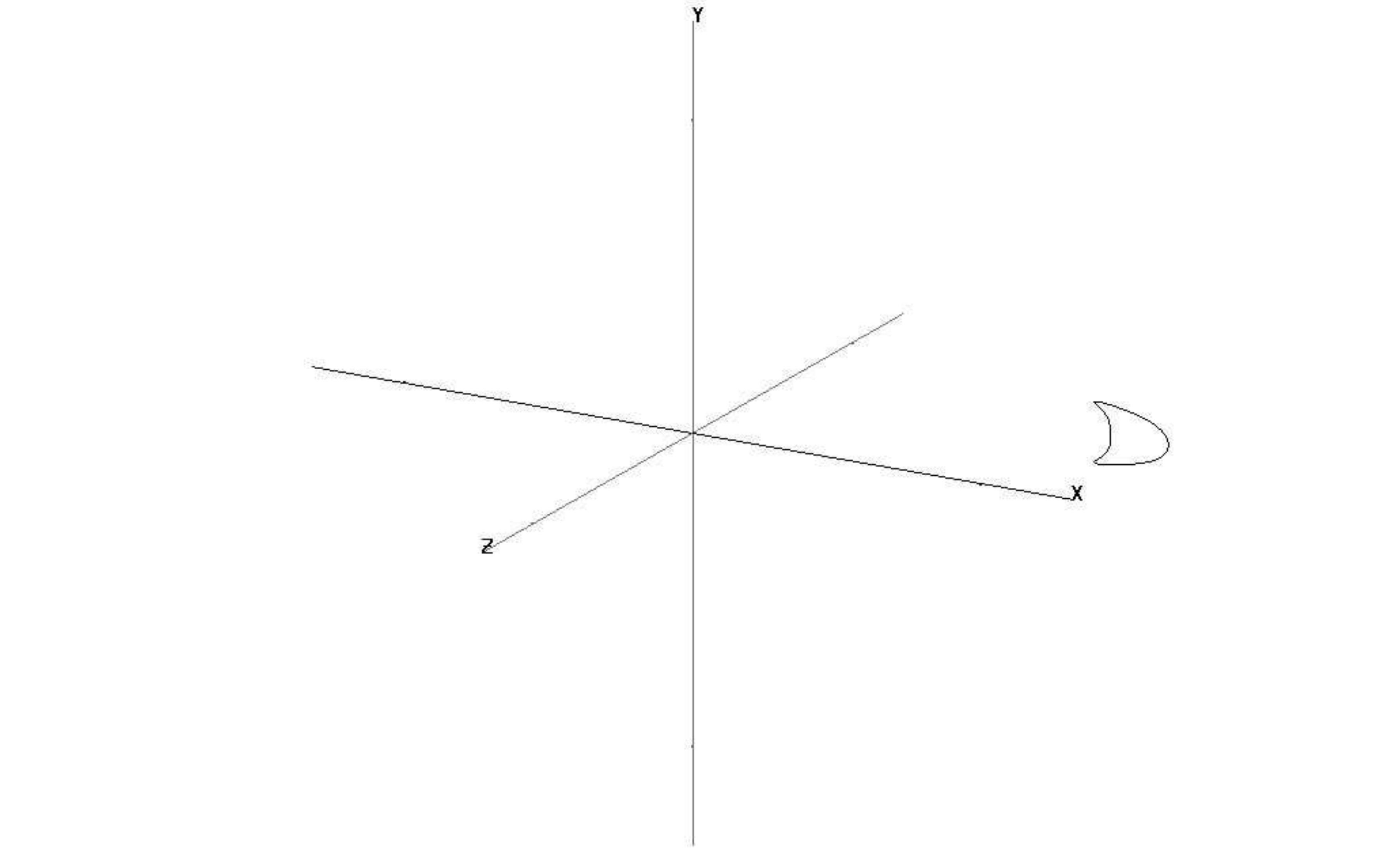}
\end{tabular}
\begin{tabular}{cc}
\includegraphics[width=7.9cm]{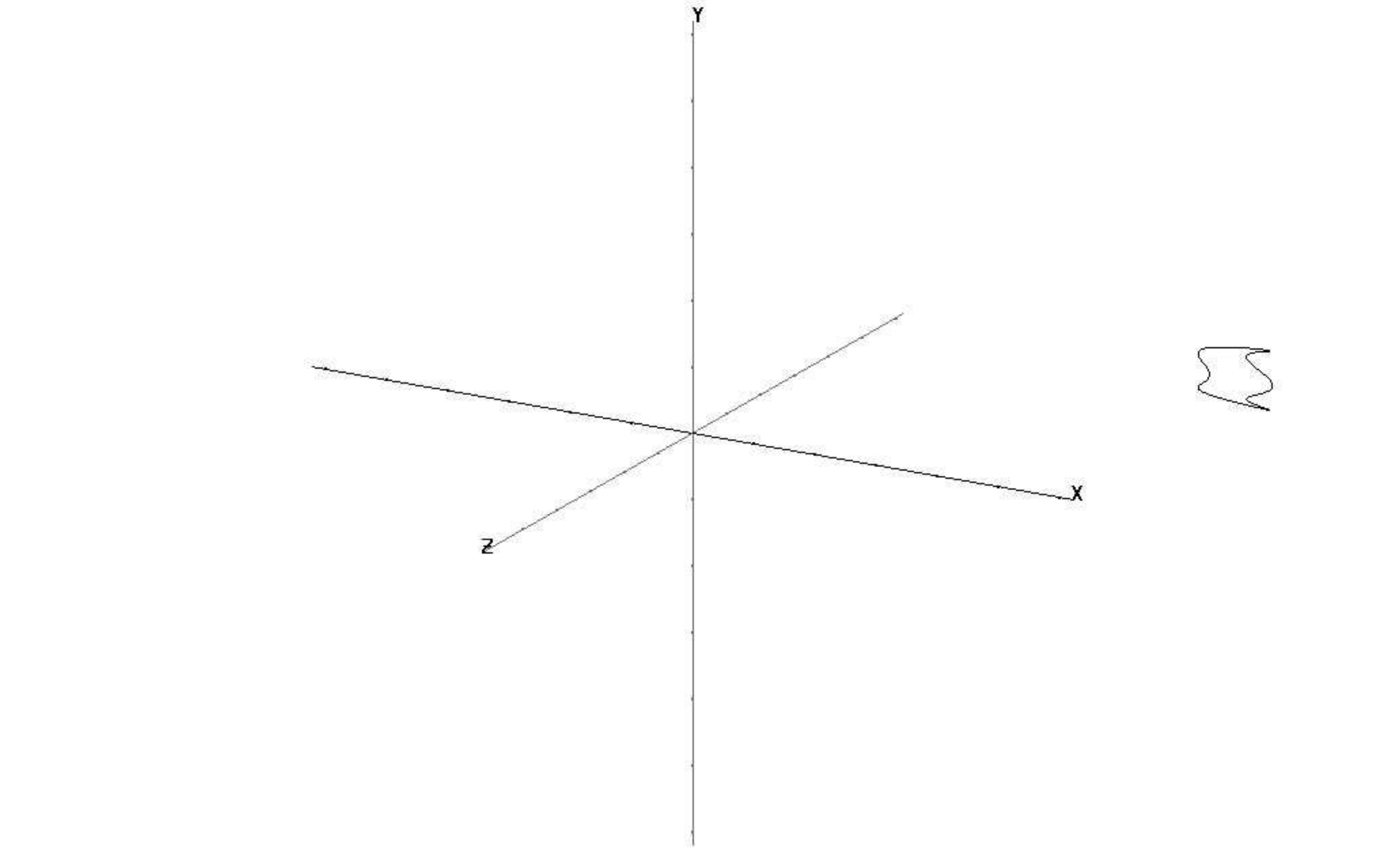}&
\mbox{$\tau=6~~~~~~~~\tau=7$}
\includegraphics[width=7.9cm]{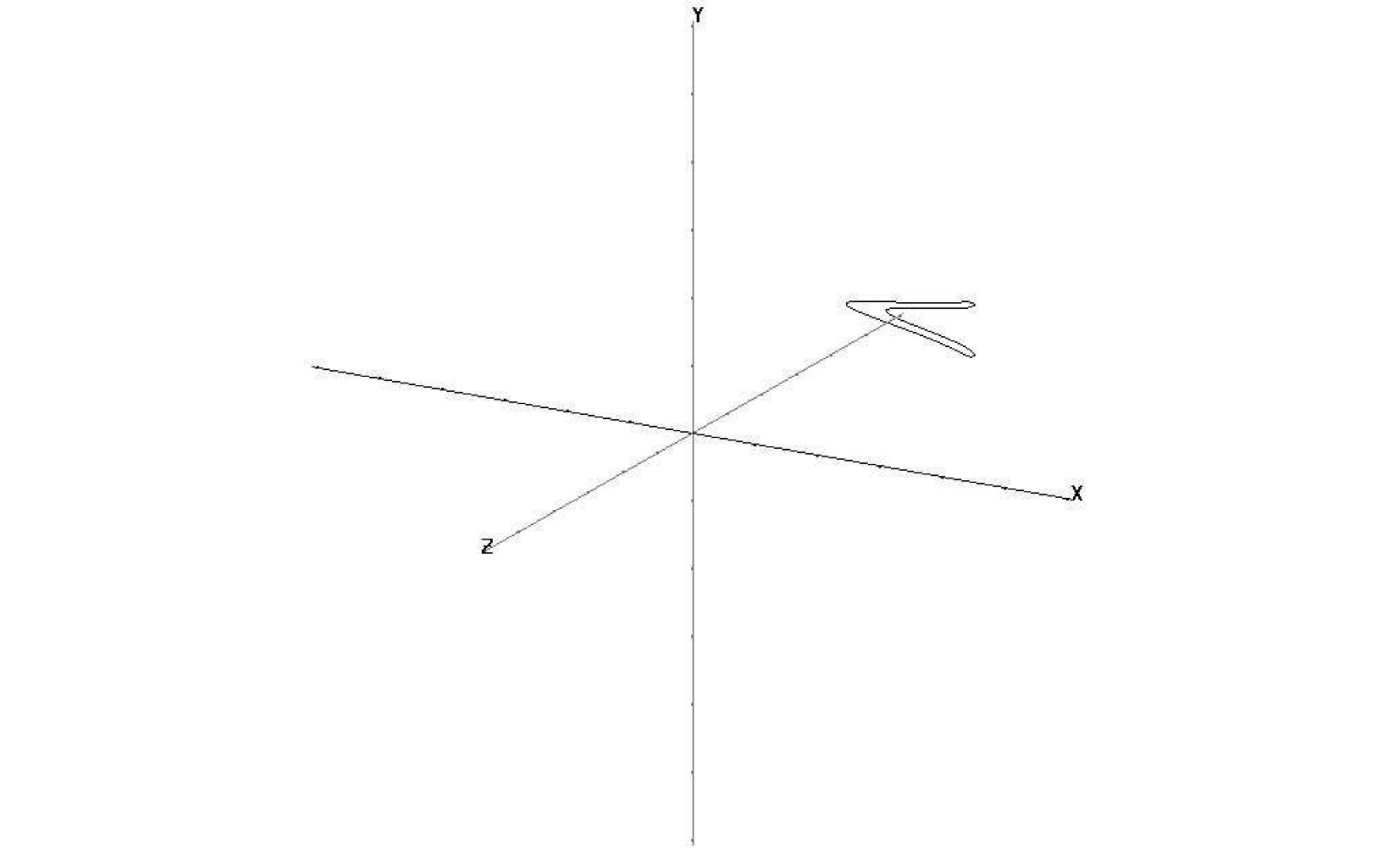}
\end{tabular}
\caption{Examples of the string's configurations in the case of radiation field. Initial conditions: $R=5$, $X_i=(0,0,20)$, $V_i=(0,0,-0.5)$.}
\label{fig2}
\end{center}
\end{figure*}
\begin{figure*}[htb]
\begin{center}
\begin{tabular}{cc}
\includegraphics[width=7.9cm]{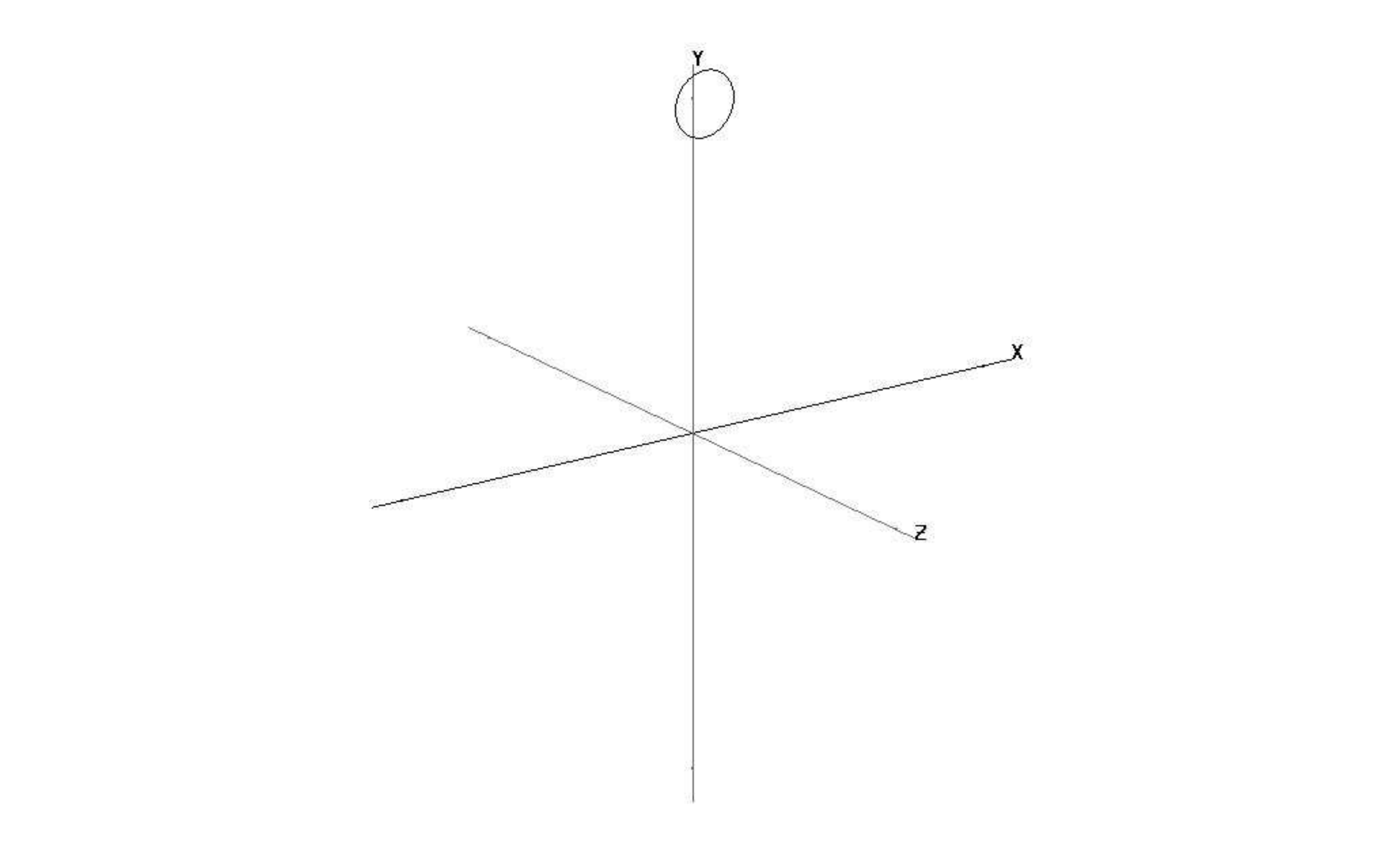}&
\mbox{$\tau=0~~~~~~~~\tau=1$}
\includegraphics[width=7.9cm]{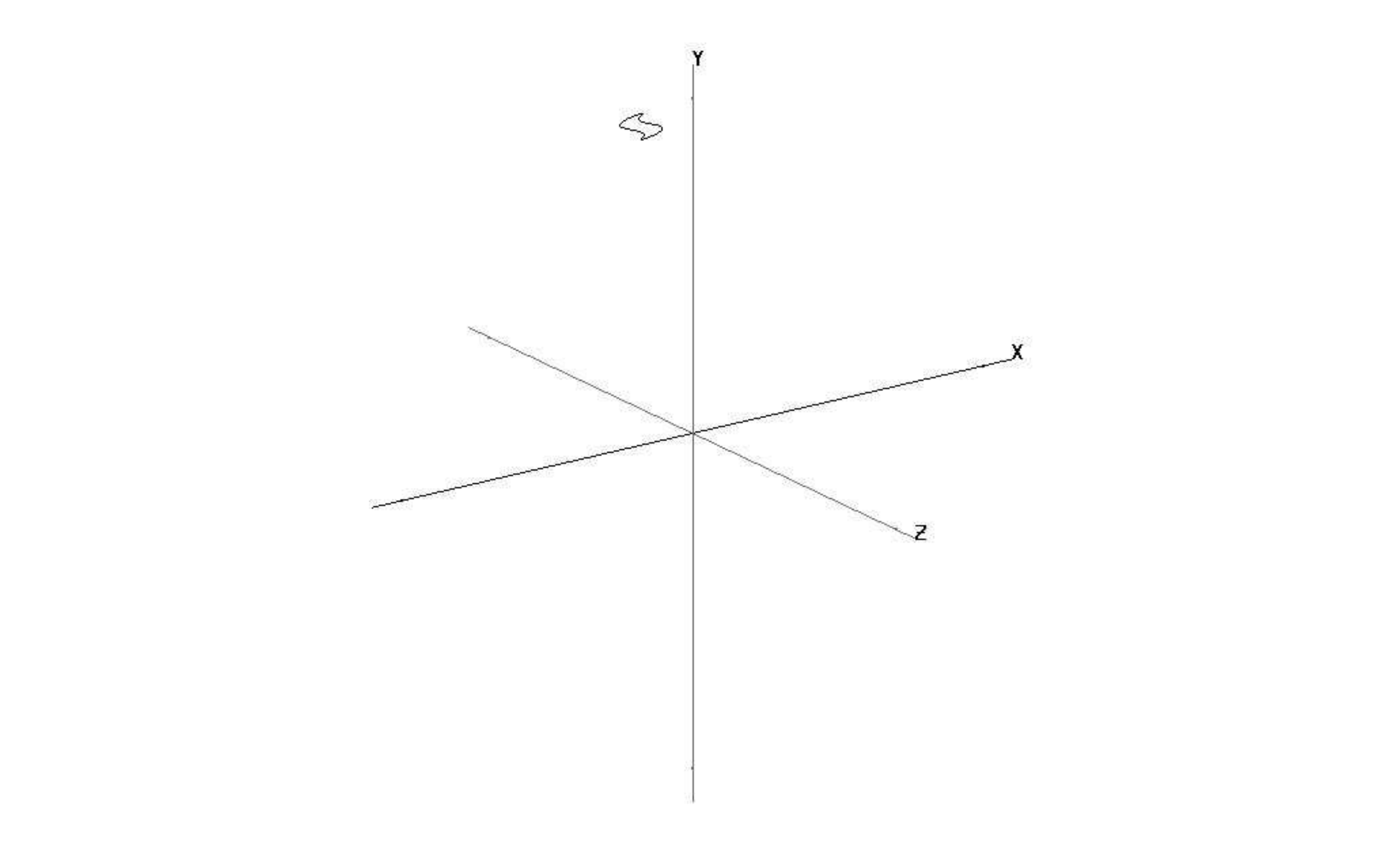}
\end{tabular}
\begin{tabular}{cc}
\includegraphics[width=7.9cm]{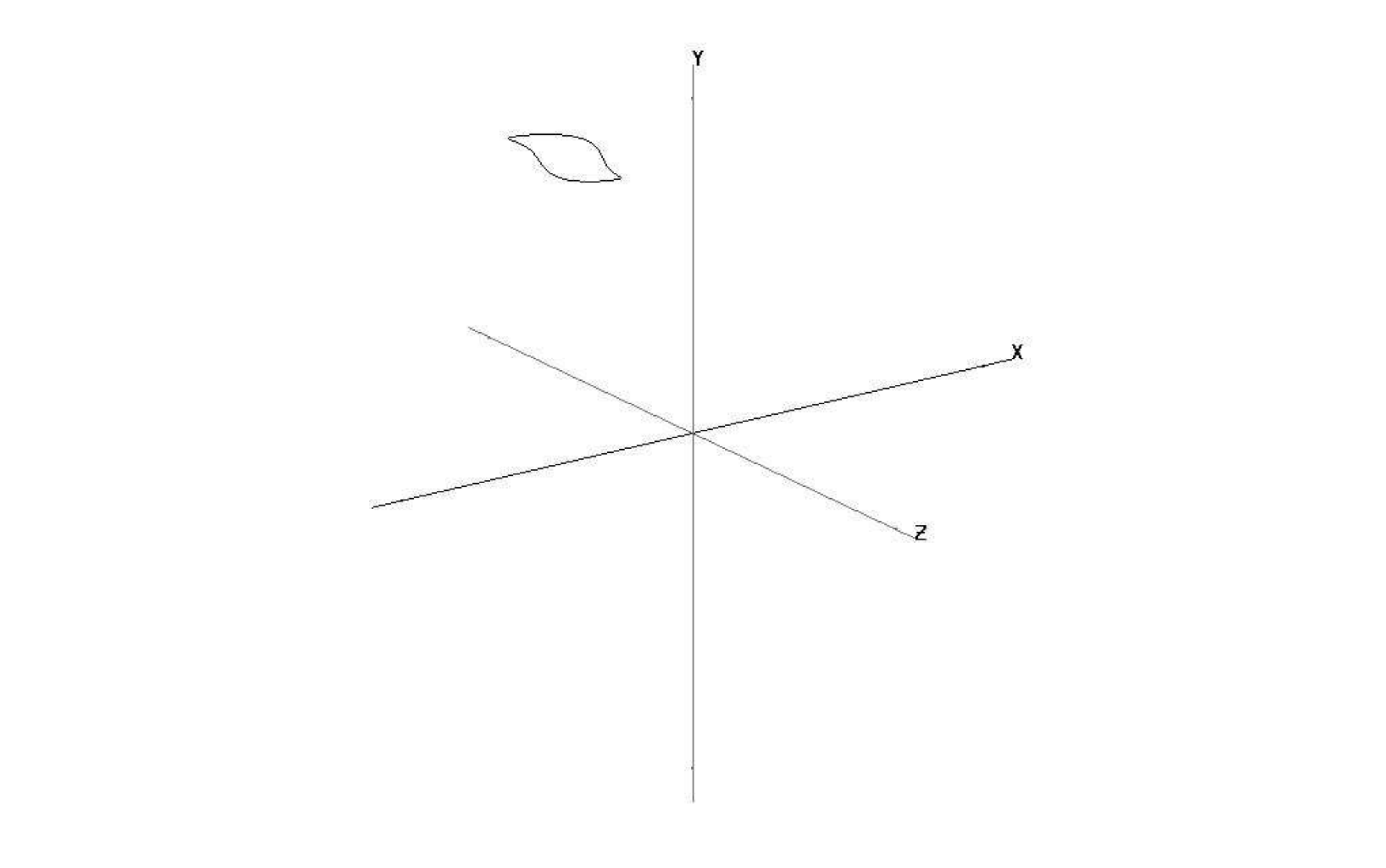}&
\mbox{$\tau=2~~~~~~~~\tau=3$}
\includegraphics[width=7.9cm]{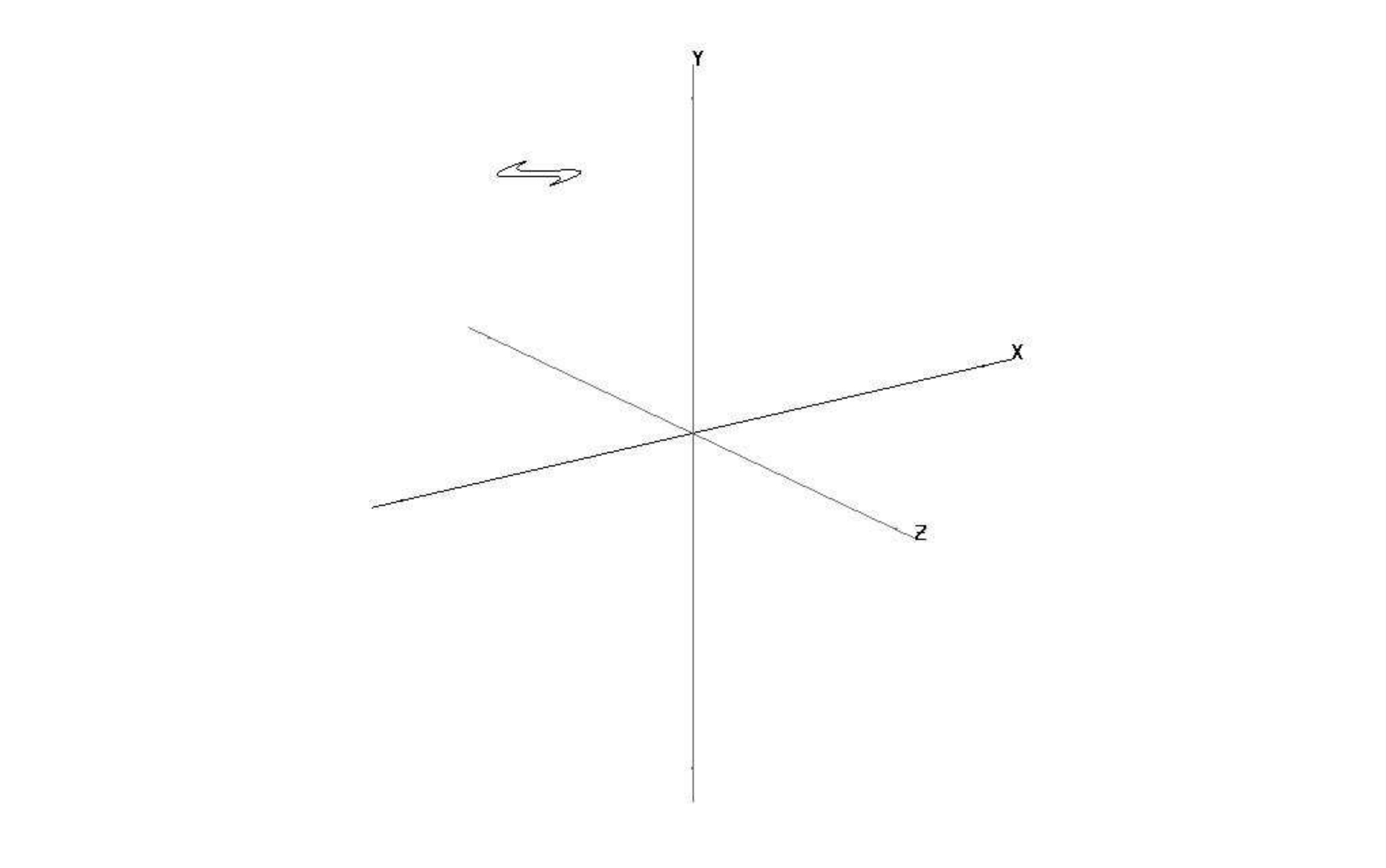}
\end{tabular}
\begin{tabular}{cc}
\includegraphics[width=7.9cm]{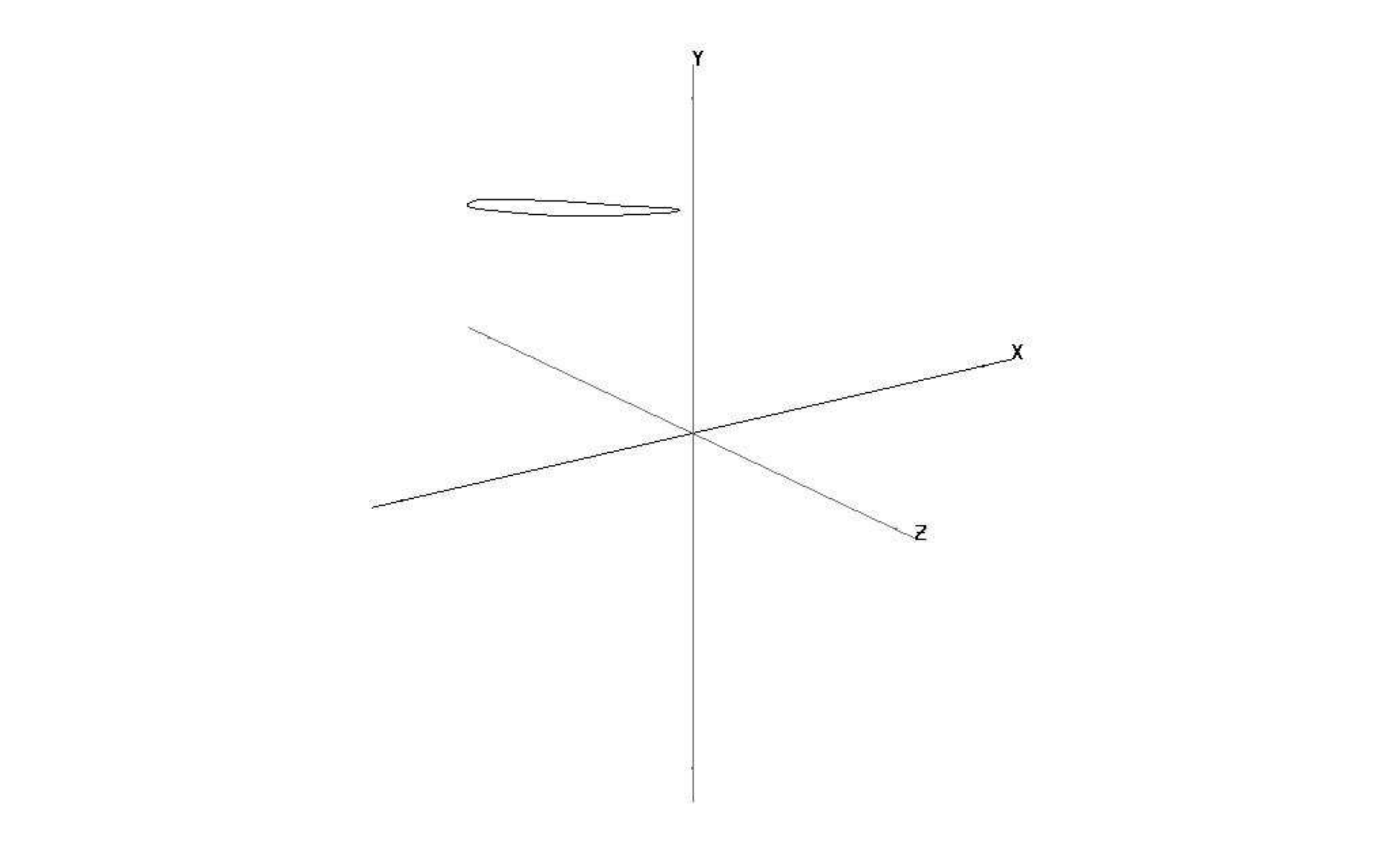}&
\mbox{$\tau=4~~~~~~~~\tau=5$}
\includegraphics[width=7.9cm]{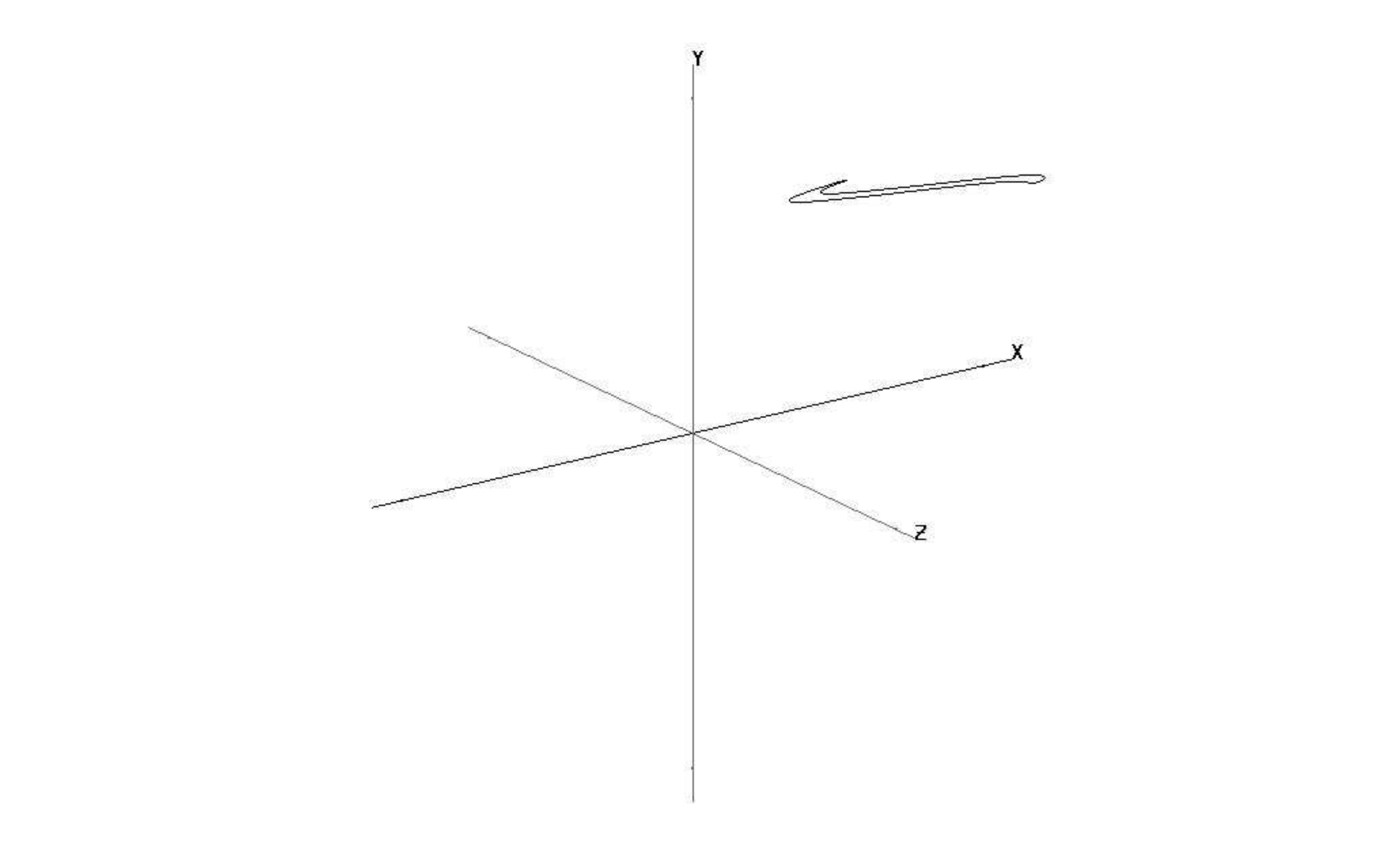}
\end{tabular}
\caption{Examples of the string's configurations in the case of strong gravitational wave. Initial conditions: $R=5$, $X_i=(0,50,3)$, $V_i=(-0.8,-0.1,0)$.}
\label{fig3}
\end{center}
\end{figure*}
\begin{gather}
\hat{x}-2x+\check{x}-\Lambda(\gamma \hat{x}+(1-2\gamma)x+\gamma \check{x})+f=0,\\
x_{m}=x^{M}(\tau_{l},\sigma_{m}),\; \hat{x}_{m}=x^{M}(\tau_{l+1},\sigma_{m}),\; \nn\\
\check{x}_{m}=x^{M}(\tau_{l-1},\sigma_{m}),\\
\Lambda x_{m}=h^2(x_{m+1}-2x_{m}+x_{m-1}),\\
f=\frac{1}{4}(\Gamma^{M}_{PQ})_{l,m}[(x^{P}_{l+1,m}-x^{P}_{l-1,m})(x^{Q}_{l+1,m}-x^{Q}_{l-1,m})\nonumber \\
-h^2(x^{P}_{l,m+1}-x^{P}_{l,m-1})(x^{Q}_{l,m+1}-x^{Q}_{l,m-1})],\\
h^2=\frac{\lambda^2(\Delta\tau)^2}{(\Delta\sigma)^2}.
\end{gather}

For all weights to be nonnegative, we must take $0\leqslant\gamma\leqslant1/2$. We also compose the difference equations for constraints (7), (9) and conditions
(14), (15):
\begin{gather}
(G_{PQ})_{l,m}[(x^{P}_{l+1,m}-x^{P}_{l-1,m})(x^{Q}_{l+1,m}-x^{Q}_{l-1,m})-{ }\nonumber \\
{ }-h^2(x^{P}_{l,m+1}-x^{P}_{l,m-1})(x^{Q}_{l,m+1}-x^{Q}_{l,m-1})]=0,\\
(G_{PQ})_{l,m}(x^{P}_{l+1,m}-x^{P}_{l-1,m})(x^{Q}_{l+1,m}-x^{Q}_{l-1,m})=0,\\
x^{N}_{0,m}=\rho^{N}(\sigma),\; x^{N}_{1,m}-x^{N}_{-1,m}=2v^{N}(\sigma_{m})\Delta\tau,\\
x^{N}_{l,-1}=x^{N}_{l,M},\; x^{N}_{l,M+1}=x^{N}_{l,0}.
\end{gather}
Equation (17) is the implicit three-layer difference equation approximating a partial differential equation with an accuracy of $O((\Delta\tau)^2+(\Delta\sigma)^2)$.
We solve Eqs. (17)--(25) using the iteration method.

We have tested Eqs. (17)--(25) against the well-known solutions for strings in the worm-hole background \cite{Zinchenko}. The test problems indicate that the proposed method is stable and the absolute error on the lattice $L\times M=1000\times 100$ does not exceed $10^{-3}$.

In figures~\eqref{fig2},~\eqref{fig3} some examples of evolution of the closed cosmic string in Peres space-time are presented. In rectangular system of coordinates the cosmic string during the initial moment of time has the form of a ring and moves as whole in some direction. And the ring plane is parallel to an anisotropy plane. At interaction with a field, the string deforms the form and it has additional components of speed. And the string remains in an initial plane and does not leave it. It is necessary to notice also, that in a case non-parallel string plane, string at the movement changes it and starts to make complex rotation as whole. Figure~\eqref{fig2} shows sequential change of position and the form of a string during of its movement for a case of a radiation field with isotropic energy-movement tensor. The world-sheet time $\tau$ numbers figures in a chronological order from initial ring configuration of cosmic string. It is well visible as initially circular string starts to be deformed strongly and gradually  involve by a field, receiving a component of speed along an axis $X$. After a while there is some kind of scattering of a string by radiation field and it changes a movement direction almost for the opposite.

Figure~\eqref{fig3} shows sequential change of position and the form of a string during of its movement for a case of a strong gravitational wave. In this case we have an opposite situation: at first the string moves contrary to wave propagation, then there is a capture of a string by a wave and string finally involve in it. It is necessary to notice also, that from a kind of the quadratic form (12) possibility of change of the signature of the metrics when a component metric tensor $g_{t t}$ changes a sign follows. Therefore at numerical calculations a sign $y^2-z^2>0$ was controlled.

\section{Conclusion}

In this paper we have considered dynamics of cosmic strings in Peres space-time, using finite-difference method. The nonlinear differential equations of motion have been written in finite differences and numerical simulation is made. It is shown that motion of the strings is very complicated. In particular, the cosmic strings make complex oscillations and deformed to form loops and cusps. Both in case of a radiation field, and in case of a strong gravitational wave, the cosmic string undergo scattering and capture or involve by the field. Also it is important to notice, that if during the initial moment of time the string is in a plane parallel to a plane of anisotropy this orientation remains at the movement. Otherwise, if during the initial moment of time the string forms some angle with this plane than at the movement orientation changes arbitrarily. In other words, in this case to oscillations of a string and its forward movement it is added also string rotation as whole.
Animation of the string movement shows characteristic "tumbles" it together with expansion and compression of a string loop and its movement as whole.

Let's make some more last remarks. Of course, in the given work we have considered idealized situation of movement of a cosmic string as test body in the given curved space-time. We have neglected back reaction, possible radiation by a string of gravitational and electromagnetic waves, finiteness of cross-section of a string and other. However, even in such simplified model, the spent numerical modeling has revealed a rich spectrum of possible types of the string movement. It is characteristic for all nonlinear problems. Therefore for qualitative understanding of dynamics of a cosmic string the given approach is quite adequate and productive.
\bibliography{ref}

\begin{thebibliography}{14}%
\makeatletter
\providecommand \@ifxundefined [1]{%
 \@ifx{#1\undefined}
}%
\providecommand \@ifnum [1]{%
 \ifnum #1\expandafter \@firstoftwo
 \else \expandafter \@secondoftwo
 \fi
}%
\providecommand \@ifx [1]{%
 \ifx #1\expandafter \@firstoftwo
 \else \expandafter \@secondoftwo
 \fi
}%
\providecommand \natexlab [1]{#1}%
\providecommand \enquote  [1]{``#1''}%
\providecommand \bibnamefont  [1]{#1}%
\providecommand \bibfnamefont [1]{#1}%
\providecommand \citenamefont [1]{#1}%
\providecommand \href@noop [0]{\@secondoftwo}%
\providecommand \href [0]{\begingroup \@sanitize@url \@href}%
\providecommand \@href[1]{\@@startlink{#1}\@@href}%
\providecommand \@@href[1]{\endgroup#1\@@endlink}%
\providecommand \@sanitize@url [0]{\catcode `\\12\catcode `\$12\catcode
  `\&12\catcode `\#12\catcode `\^12\catcode `\_12\catcode `\%12\relax}%
\providecommand \@@startlink[1]{}%
\providecommand \@@endlink[0]{}%
\providecommand \url  [0]{\begingroup\@sanitize@url \@url }%
\providecommand \@url [1]{\endgroup\@href {#1}{\urlprefix }}%
\providecommand \urlprefix  [0]{URL }%
\providecommand \Eprint [0]{\href }%
\providecommand \doibase [0]{http://dx.doi.org/}%
\providecommand \selectlanguage [0]{\@gobble}%
\providecommand \bibinfo  [0]{\@secondoftwo}%
\providecommand \bibfield  [0]{\@secondoftwo}%
\providecommand \translation [1]{[#1]}%
\providecommand \BibitemOpen [0]{}%
\providecommand \bibitemStop [0]{}%
\providecommand \bibitemNoStop [0]{.\EOS\space}%
\providecommand \EOS [0]{\spacefactor3000\relax}%
\providecommand \BibitemShut  [1]{\csname bibitem#1\endcsname}%
\let\auto@bib@innerbib\@empty
\bibitem{Peebles}
P.J.E. Peebles, {\it Principles of  Physical Cosmology} (Princeton University Press, 1994).
\bibitem{Vilenkin}
A. Vilenkin and E.P.S. Shellard, {\it Cosmic Strings and Other Topological Defects} (Cambridge University Press, 1994).
\bibitem{de_Vega1}
H.J. de Vega, ``Strings in Curved Space-Times'', (1993), arXiv:hep-th/9302052 [hep-th].
\bibitem{de_Vega2}
H.J. de Vega and I.L. Egusquiza, ``String in Cosmological and Black Hole Backgrounds: Ring Solutions'', (1993), arXiv:hep-th/9309016 [hep-th].
\bibitem{Larsen1}
A.L. Larsen, ``Circular Strings in de Sitter Space-Time'', (1994), arXiv:hep-th/9408026 [hep-th].
\bibitem{Larsen2}
A.L. Larsen and N. Sanchez, ``Strings and Multi-Strings in Black Hole and Cosmological Space-Times'', (1995), arXiv:hep-th/9504007 [hep-th].
\bibitem{Anderson}
M.R. Anderson, {\it The Mathematical Theory of Cosmic Strings} (Institute of Physics Publishing Bristol and Philadelphia, 2003).
\bibitem{Larsen3}
A.V. Frolov and A.L. Larsen, ``Chaotic Scattering and Capture of Strings by Black Hole'', (1999), arXiv:gr-qc/9908039v2 [gr-qc].
\bibitem{Zhovtan}
A.V. Zhovtan and S.N. Roshchupkin, Ukr. J. Phys. v.50, p.639, (2005).
\bibitem{Lelyakov}
A.P. Lelyakov and S.N. Roshchupkin, Acta Phys. Polonica B, v.33, p.593, (2002).
\bibitem{Zheltukhin}
A.A. Zheltukhin and S.N. Roshchupkin, Sov. J. Teor. Mat. Fiz. 111, p.402, (1997).
\bibitem{Peres}
A. Peres, Phys. Rev. v.118, p.1105, (1960).
\bibitem{Ivanov}
G.G. Ivanov, Izv. Vuzov. Math. v.9, p.64, (1985).
\bibitem{Zinchenko}
E.N. Zinchenko and S.N. Roshchupkin, Ukr. J. Phys. v.47, p.113, (2002).
\end{thebibliography}%
\end{document}